\documentclass{mem}
\usepackage{natbib}\usepackage{txfonts}\usepackage{balance}
\usepackage{graphicx}
\usepackage[a4paper]{hyperref}
\idline{75}{282}

\begin{document}
\def\teff{$T\rm_{eff }$}
\def\kms{$\mathrm {km s}^{-1}$}
\def\msun{\,{\rm M_\odot}}
\newcommand{\etal}{et al.\ }
\title{
HARDENING IN A TIME--EVOLVING STELLAR BACKGROUND: 
HYPER--VELOCITY STARS, ORBITAL DECAY AND PREDICTION FOR {\it LISA}
}

   \subtitle{}

\author{
F. \,Haardt\inst{1}, A. \,Sesana\inst{1}, P. \,Madau\inst{2} 
          }

  \offprints{F. Haardt}

\institute{
Universit\'a degli Studi dell'Insubria --
Dipartimento di Fisica e Matematica, Como, Italy
\and
Department of Astronomy --
University of California, Santa Cruz (CA), USA
}

\authorrunning{F. Haardt}

\titlerunning{Black Hole Binaries}

\abstract{We study the long-term evolution of massive black hole binaries (MBHBs) 
at the centers of galaxies using detailed full three-body scattering experiments.
Stars, drawn from a distribution unbound to the binary, are ejected by the gravitational slingshot. We 
quantify the effect of secondary slingshots -- stars returning on small impact parameter orbits 
to have a second super-elastic scattering with the MBHB -- on binary separation.
Even in the absence of two-body relaxation or gas dynamical processes,
very unequal mass binaries of mass $M=10^7\,\msun$ can shrink to the gravitational 
wave emission regime in less than a Hubble time, and are therefore a target
for the planned {\it Laser Interferometer Space Antenna (LISA)}. 
Three-body interactions create a subpopulation of hypervelocity stars on nearly
radial, corotating orbits, with a spatial distribution that is initially highly flattened 
in the inspiral plane of the MBHB, but becomes more isotropic with decreasing binary 
separation.  The mass ejected is $\gtrsim 0.7$ times the binary reduced mass, and most of the 
stars are ejected in an initial burst lasting much less than a bulge crossing time.

}
\maketitle{}

\section{Introduction}

It is now widely accepted that the formation and evolution of galaxies and
massive black holes (MBHs) are strongly linked: MBHs are ubiquitous in the nuclei of nearby
galaxies, and a tight correlation is observed between hole mass and the stellar mass of
the surrounding spheroid or bulge (e.g. Magorrian et al. 1998; Gebhardt et al. 2000;
Ferrarese \& Merritt 2000; Haring \& Rix 2004).
If MBHs were also common in the past (as implied by the notion that many distant galaxies
harbor active nuclei for a short period of their life), and if their host
galaxies experience multiple mergers during their lifetime, as dictated by
popular cold dark matter (CDM) hierarchical cosmologies, then close MBH binaries (MBHBs)
will inevitably form in large numbers during cosmic history. A MBHB model for the
observed bending
and apparent precession of radio jets from active galactic nuclei was first proposed
by Begelman, Blandford, \& Rees (1980). The coalescence of two spinning black holes in a radio
galaxy may cause a sudden reorientation of the jet direction, perhaps
leading to the so-called ``winged'' or ``X-type'' radio sources
(Merritt \& Ekers 2002). Recently, observations with the {\it Chandra} satellite
have revealed two active MBHs in the nucleus of NGC 6240 (Komossa et al. 2003), and a MBHB
is inferred in the radio core of 3C 66B (Sudou et al. 2003).

BH pairs that are able to coalesce in less than a Hubble time will
give origin to the loudest gravitational wave (GW) events in the universe.
In particular, a low-frequency space interferometer like the planned {\it Laser
Interferometer Space Antenna (LISA)} is expected to have the sensitivity to detect
nearly all MBHBs
in the mass range $10^4-10^7\, \msun$ that happen to merge at any redshift during the
mission operation phase (Sesana et al. 2005). The coalescence rate of such ``{\it LISA}
MBHBs'' depends, however, on the efficiency with which stellar and gas dynamical
processes can drive wide pairs to the GW emission stage.

Following the merger of two halo$+$MBH systems of comparable
mass (``major mergers''), it is understood that dynamical friction will drag
in the satellite halo (and its MBH) toward the center of the more massive
progenitor (see, e.g., the recent numerical simulations by Kazantzidis et al. 2005):
this will lead to the formation of
a bound MBH binary in the violently relaxed core of the newly merged stellar system.
As the binary separation decays, the effectiveness of dynamical friction slowly declines
because distant stars perturb the binary's center of mass but not its semi-major axis
(Begelman, Blandford, \& Rees 1980). The bound pair then hardens by capturing
stars passing  in its immediate vicinity and ejecting them at much higher
velocities (gravitational slingshot).
It is this phase that is considered the bottleneck of a MBHB's path to
coalescence, as there is a finite supply of stars on intersecting orbits
and the binary may ``hung up'' before the back-reaction from GW emission becomes
important. This has become known as the ``final parsec problem'' (Milosavljevic \& Merrit 2003).

Recently, in collaboration with Piero Madau and Monica Colpi, the group in Como  
started a study of MBHB dynamics from two different perspectives, 
namely, the role of gas in the dynamical friction regime (the 100-to-1 pc decay), and the role 
of 3--body encounters with bulge stars. The interested reader can find exahustive descriptions 
of computational tools, and detailed discussions of the results in Dotti, Colpi \& Haardt (2006), and 
in Sesana, Haardt \& Madau (2006).  We have tried to answer a number of questions, i.e.,

i) Is stellar slingshot able to drive MBHBs to the GW regime in an Hubble time?
What is the role of different mass ratios and eccentricities in setting the orbital decay time scale? 

ii) What are the cinematical properties of the stellar population after the interaction 
with the MBHB? Hyper velocity stars can be an observable large scale signature of a past MBHB merging.

iii) How do eccentric orbits evolve in a gaseous circum nuclear disk? 
Do they become circular?  This
issue may be relevant in establishing the initial conditions for the
braking of the binary due to the slingshot mechanism and/or
gaseous gravitational torque. 

iv) During the sinking process, do the BHs collect substantial
amounts of gas?  This is a query related to the potential activity of
a MBH during a merger and its detectability across the entire
dynamical evolution. This issue is related to the search of EM 
counterparts to GW signals.

In the following, I briefly review our main results concerning the hardening 
in a time evolving stellar background, i.e., I will discuss our answers to questions i) and ii) above.

\section{Scattering of unbound stars}

\subsection{Hyper-velocity stars}
Assume a MBHB of mass $M=M_1+M_2=M_1(1+q)$ ($M_2<M_1$), and initial eccentricty $e$, embedded in 
a stellar cusp of mass $M_B$ described by Maxwellian distribution function, with mean velocity $\sigma$.
Quinlan (1996) showed that the hardening rate is relevant, and stays almost constant, for 
separation smaller than 
\begin{equation}
a_h=\frac{GM_2}{4\sigma^2},
\label{eq:ah}
\end{equation}
known as the ``hardening'' radius of the MBHB. Here, it is important to notice that, by measuring binary 
separations in units of $a_h$, scattering experiments results are independent on any pre--asigned value
of $\sigma$.

From the point of view of stars, repeated singshots results in a net heating of the distribution function, 
with a consequent erosion of the inner cusp. 
Though the overall energy budget is modest, as only a small fraction 
($\lesssim 1$\%) of the stars are in the loss cone at $t=0$, nevertheless a substantial population 
of high velocity (often known as {\it suprathermal}) stars is produced. We define these stars 
by the condition $v>v_{\rm esc}$, where the escape velocity $v_{\rm esc}$ is taken as   
\begin{equation}
v_{\rm esc} = 2\sigma \sqrt{\ln(M_B/M)}\simeq 5 \sigma.
\label{eq:vesc}
\end{equation}
Equation \ref{eq:vesc} gives the escape velocity from the MBHB influence radius $r_i=GM/2\sigma^2$. 
The second equality comes from the adopted relation $M=0.002\,M_B$.

\begin{figure*}[t!]
\resizebox{\hsize}{!}{\includegraphics[clip=true]{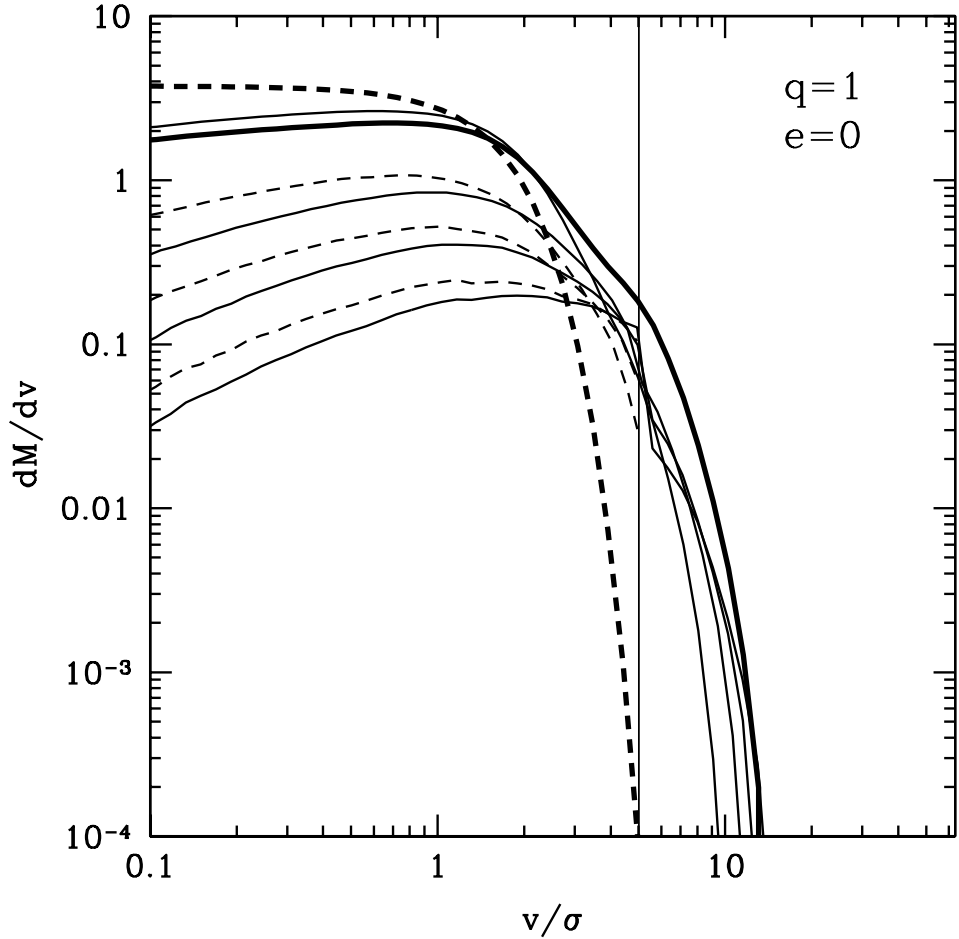}}
\caption{\footnotesize Stellar velocity distribution (normalized to the total mass of interacting stars in units of the total 
binary mass $M$) for an equal mass, circular binary, at different stages of the hardening. 
The vertical line marks $v_{\rm esc}\simeq 5 \sigma$.
{\it Dashed lines:} from top to bottom, distribution of stars in the shrinking  
loss cone before the 1st, 2nd, 3rd and 4th iteration. For clarity, the initial loss cone distribution is 
marked with a thicker line. {\it Thin solid lines:} from top to bottom, distribution of stars which have 
received 1, 2, 3 and 4 kicks. 
{\it Thick solid line:} final stellar velocity distribution after loss cone depletion is completed.}    
\label{fig:dMdv}
\end{figure*}

An example of the effects of the slingshot mechanism on the stellar population is shown in 
Figure~\ref{fig:dMdv}, where the initial ($t=0$) velocity distribution (stellar mass per unit velocity) of 
interacting stars is compared to that after loss cone depletion, for an equal mass, circular binary. 
It is clear that after the interaction
with the binary, a large subset of kicked stars still lies in the (reduced) loss cone 
of the shrunk MBHB, i.e., they are potentially avaliable for
further interactions. We term stars that undergo secondary slingshots ``returning''.  
Note that a large fraction of 
returning stars have velocities just below $v_{\rm esc}$. The fraction of 
secondary slingshots is relevant because the interaction with the MBHB, on one 
hand, increases the star velocity,
while, on the other, moves the kicked stars on nearly radial orbits.

Our calculations show that the high velocity tail of the distribution depends on the 
MBHB eccentricity, although the effect of changing $e$ is small for small values of $q$. 
In this case, fewer stars are kicked out compared to the case $q\lesssim 1$, but, on average, 
at higher velocities. In general, both a 
small mass ratio and a high eccentricity increase the tail of high velocity stars. 
Integrating the curves over velocity 
gives the mass of interacting stars, which turns out to be $\simeq 2M$ for equal mass binaries, 
$\simeq 1.2M$ for $q=1/3$ ($e=0.3$), and $\simeq 0.6M$ for $q=1/27$ ($e=0.3$). 
   
\begin{figure*}[t!]
\resizebox{\hsize}{!}{\includegraphics[clip=true]{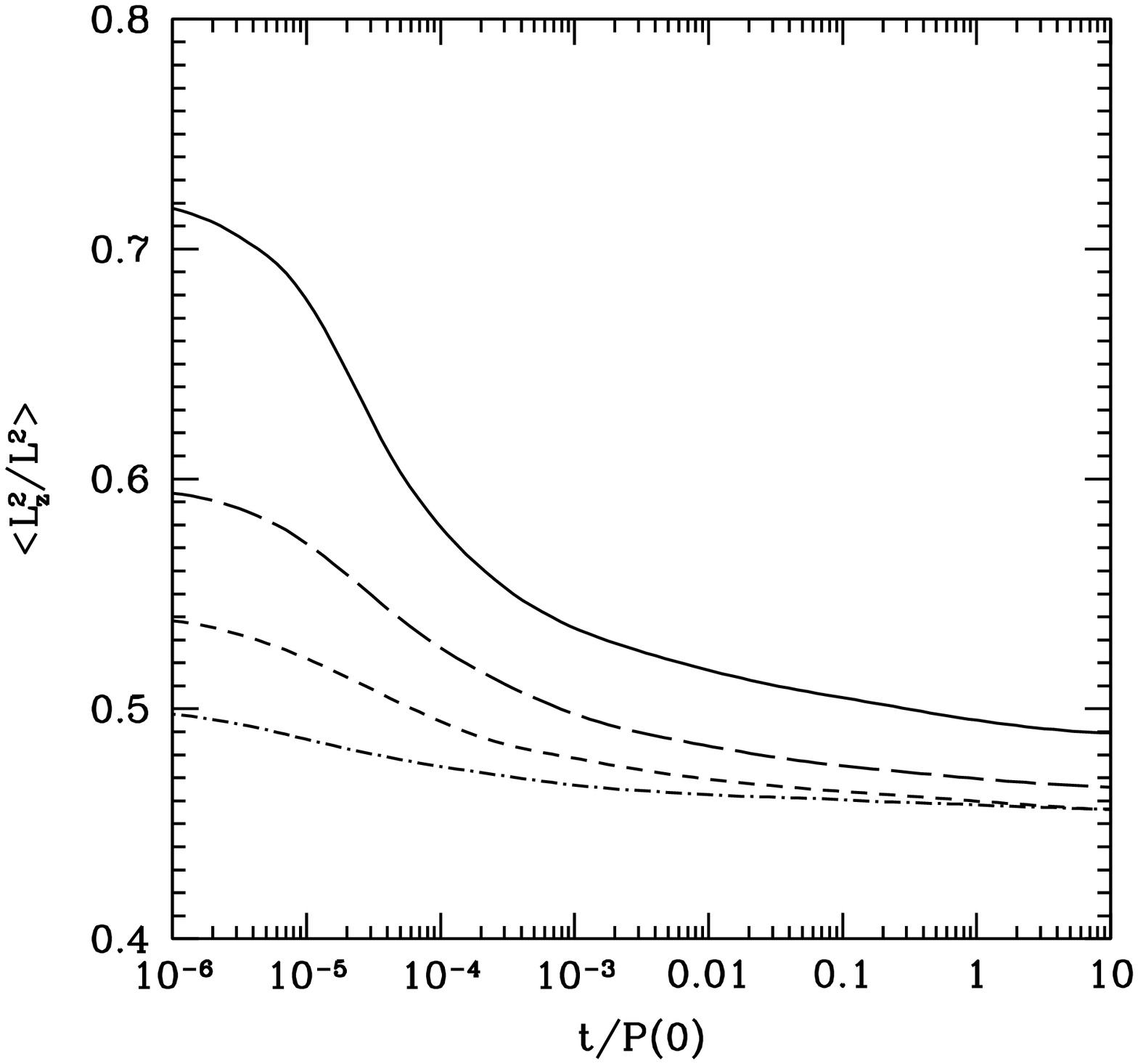}}
\caption{\footnotesize Time evolution of angular momentum of ejected stars, in terms of 
$L_{z\star}^2/L_{\star}^2$, for a circular binary, and, from bottom to top,  
for $q=$1/27, 1/9, 1/3 and 1 }
\label{fig:Lztime}
\end{figure*}

As already mentioned, not only the velocity, but also the angular momentum of scattered stars is modified 
by the slingshot process. One of the more interesting effect is the flattening of the distribution of scattered stars 
into the orbital plane of the binary. Moreover, scattered stars preferentially co-rotate with the binary. 
As a general trend, we can conclude that stars that acquire high velocities after the interaction, 
are preferentially moved into the binary orbital plane, on nearly radial, corotating orbits. 

Recently, Levin (2005) pointed out that the anisotropy of the ejected star distribution is a decreasing function of 
time. We find a qualitatively similar result in our scattering experiments, though the phyiscal context is quite different. 
In figure \ref{fig:Lztime}, the time evolution of $L_{z\star}^2/L_{\star}^2$ for the ejected stars is shown, for 
different values of $q$ (assuming $e=0$). As the binary orbit decays, ejected stars are more and more isotropic. 
In fact, for small separations, $V_c\gg v_{\rm esc}$, and even weak interactions can lead to final 
star velocities $\gtrsim v_{\rm esc}$. The effect tends to be suppressed for small $q$, simply because for small 
mass ratios $V_c \sim v_{\rm esc}$ already at $a=a_h$.

\subsection{Mass ejection and coalescence}

Integrating the mass ejection rate $J$ along the shrinking orbit, we can derive 
the ejected mass $M_{\rm ej}$. Note that, while $J$ is indipendent on the 
time evolution of the orbit, $M_{\rm ej}$ is an outcome of the hybrid model of the loss cone/orbit 
evolution.
 
\begin{figure*}[t!]
\resizebox{\hsize}{!}{\includegraphics[clip=true]{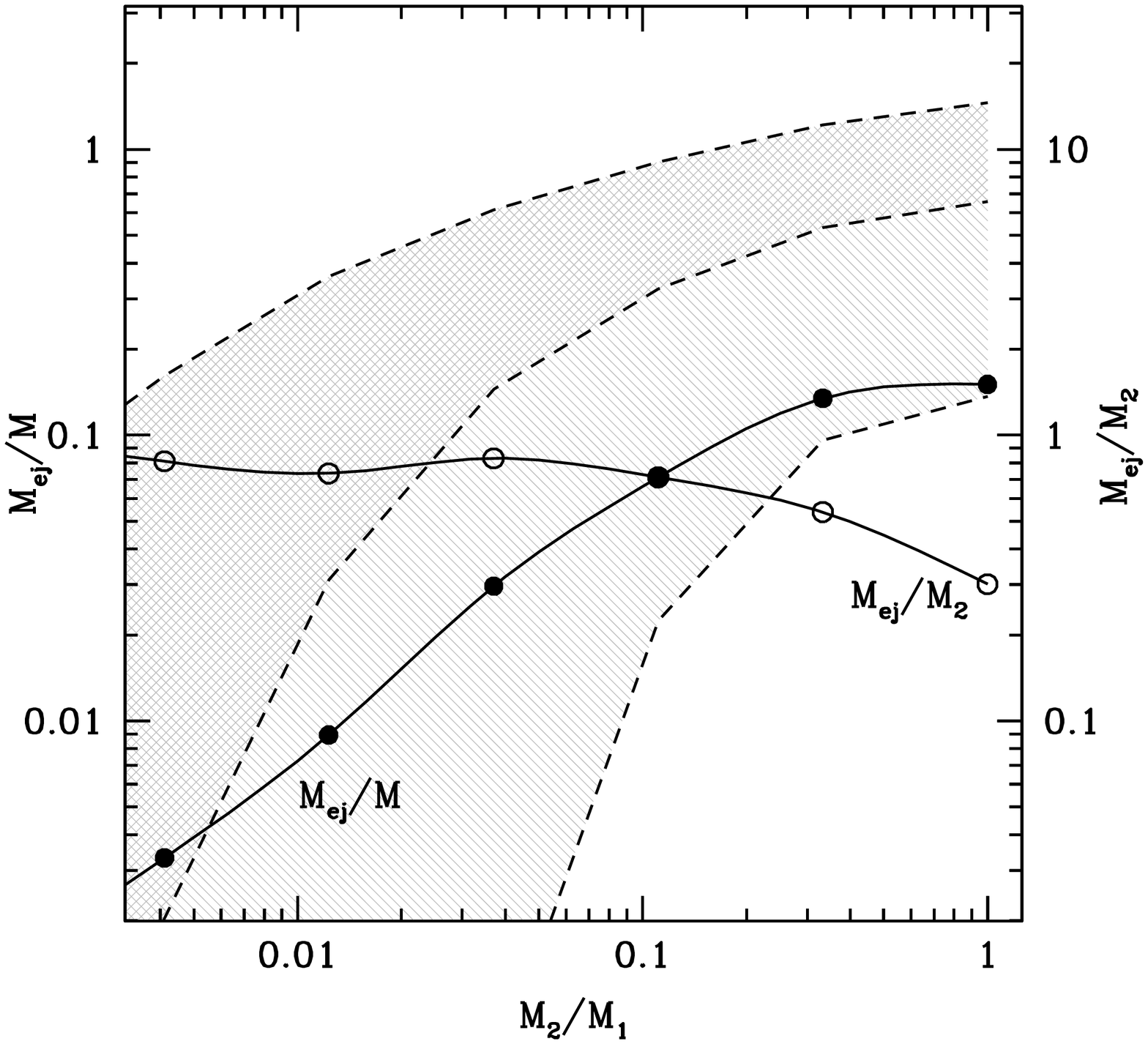}}
\caption{\footnotesize Ejected stellar mass $M_{\rm ej}$ normalized to the total binary mass $M$ (left scale, solid points), and to the 
mass of the lighter binary member $M_2$ (right scale, empty points), as a function of the binary mass ratio. 
The curves are polynomial interpolations. Note that $M_{\rm ej}/M$ and $M_{\rm ej}/M_2$ does not depend on the absolute value 
of $M$, and are nearly independent on $e$.
The upper, dark shaded area defines the mass (normalized to $M$) a $M=10^6 \msun$ binary needs to eject to reach a final separation 
$a_f$ such that $t_{\rm GW}=1$ Gyr. Upper and lower limits to the area are for $e=0$ and $e=0.9$, respectively.  
The lower, light shaded area is analogous, but for a $M=10^9 \msun$ binary. Incidentally,  
the $e=0.9$ limit for the light binary practically coincides with the $e=0$ limit of the heavy one.}
\label{fig:M_ej}
\end{figure*}

In Figure~\ref{fig:M_ej} results are given in  terms of $M_{\rm ej}$ normalized 
to $M$ (left scale) and to $M_2$ (right scale), as a function of $q$. Our results show that   
$M_{\rm ej}\sim 0.7 \mu$, i.e., $M_{\rm ej}/M \sim 0.7 q/(1+q)^2$ and $M_{\rm ej}/M_2 \sim 0.7/(1+q)$. 
Note that $M_{\rm ej}/M$ (and $M_{\rm ej}/M_2$) is independent on the absolute value of $M$. 
We also checked that the value of the eccentricity is basically irrelevant for what concerns $M_{\rm ej}$. 
The bulk of the mass 
is ejected in a initial burst, though the effect is reduced for small $q$. The burst is due to the ejection of those stars 
present in the geometrical loss cone when the binary first becomes hard. For small $q$, mass ejection is 
already relevant at $a\simeq a_h$, as in this case $V_c$ is always $\gtrsim v_{\rm esc}$.  

We may ask now if the amount of ejected mass is sufficient to shrink the MBHB orbit down to the GW--dominated regime.
The time needed to reach coalescence emitting gravitational waves for a binary of mass $M$, initial eccentricity $e$, 
and separation $a$, is computed as in Peters (1964).

\begin{figure*}[t!]
\resizebox{\hsize}{!}{\includegraphics[clip=true]{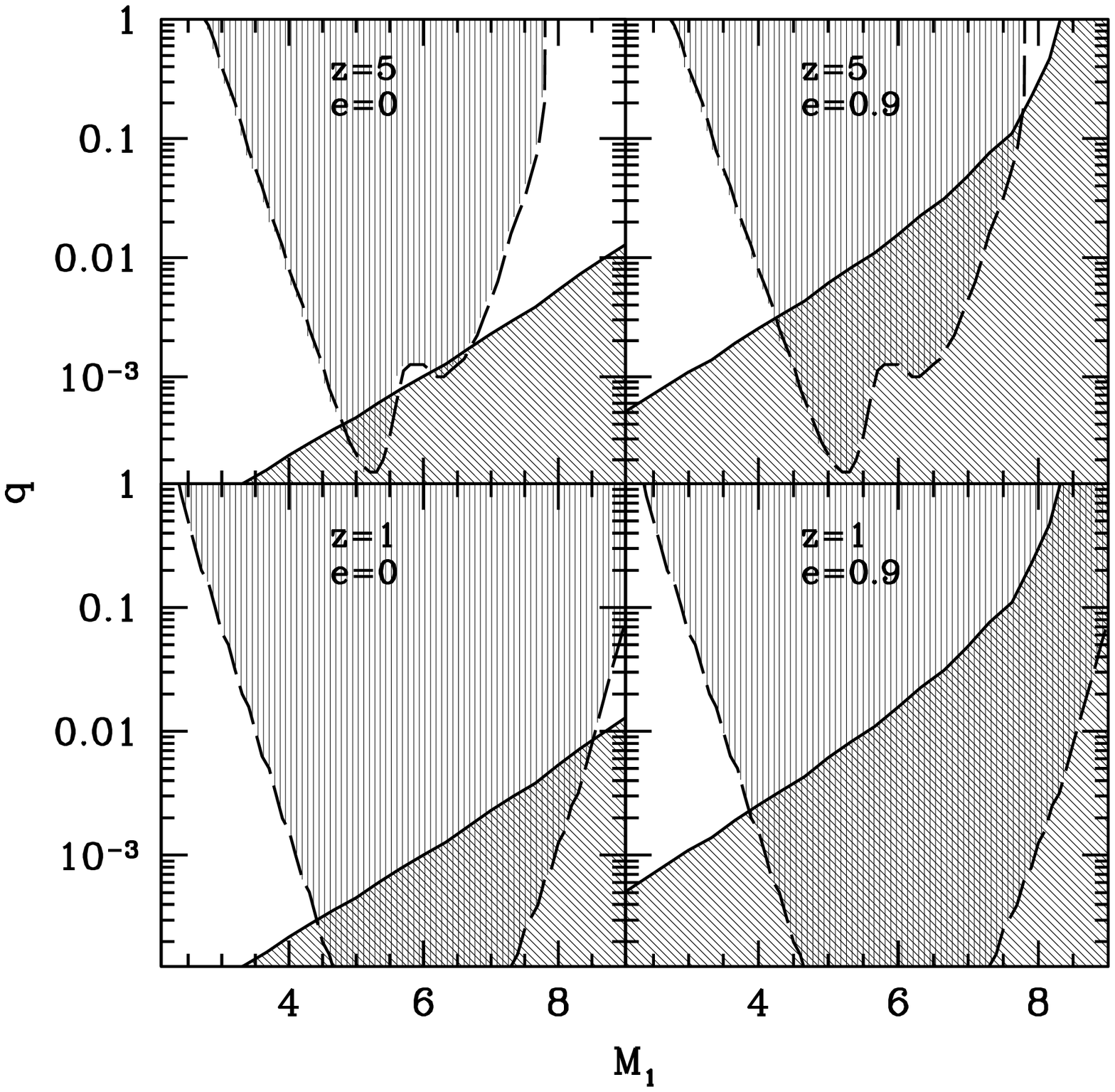}}
\caption{\footnotesize In the plane $M_1-q$, the vertical shaded area limits {\it LISA} potential targets with $SNR>5$. The diagonal shaded area on the 
lower right corner marks binaries that are going to coalesce within 5 Gyrs after loss cone depletion. In each panel, the reference redshift 
and eccentricity of the MBHBs are labelled.}
\label{fig:LISA}
\end{figure*}

Given that, it is straightforward to estimate the mass that needs to be ejected to have 
$t_{\rm GW}(a_f)=1$ Gyr, i.e., the mass ejection 
needed to reach a final orbital separation at which GW emission leads to coalescence within 1 Gyr. 
Results are shown in Figure~\ref{fig:M_ej}. The upper, dark shaded area defines such mass  
(in units of $M$) for a $10^6 \msun$ MBHB. The upper area limit is in case of a circular binary, the lower limit for $e=0.9$. The 
lower, light shaded area defines the same ``mass that needs to be ejected'' for a $10^9 \msun$ MBHB. Again, upper and lower limits are 
in the case $e=0$ and $e=0.9$, respectively. Incidentally,  
the $e=0.9$ limit of the light MBHB practically coincides with the $e=0$ limit of the heavy one.
Comparing these estimates with our $M_{\rm ej}/M$ curve (which, we recall, does not depend upon $M$ nor $e$), 
we can conclude that, if stellar slingshot is the only mechanism driving MBHBs to $a_f$, circular binaries are not going to coalesce within 
1 Gyr after they first become hard. In the case of $M=10^6 \msun$, only extreme unequal mass, highly eccentric MBHB will be led to coalescence 
by GW emission (and hence, would be detectable by {\it LISA}). Higher masses are favored, but, in any case, still 
relevant eccentricities and/or small mass ratios are needed. Figure~\ref{fig:LISA} shows the binaries (at two different reference redshifts) 
in the $M_1-q$ plane which would be 
resolved by {\it LISA} with $SNR>5$, and the binaries (in the same plane) which are going to coalesce within $5$ Gyrs after loss cone depletion 
is completed. The overlap of the two areas gives the MBHBs which are potential targets for {\it LISA}. As expected, 
the eccentricity is a crucial parameter, and it is clear how, even considering only 
slingshot of unbound stars as orbital decay driver, very unequal mass binaries are ideal {\it LISA} targets. 

\section{Conclusions}
We have made a first attempt of modeling the time dependent dynamical evolution of MBH pairs in stellar backgrounds 
using an hybrid approach, i.e., using results of three-body scattering experiment to follow the time evolution of 
the stellar distribution. Despite several approximations and limitations, we have obtained a number of results 
we briefly summarize in the following. 

A net outcome of the MBHB-stellar interaction is a significative mass ejection, that we found scaling as 
$M_{\rm ej}\simeq 0.7 \mu$, nearly independent on the binary eccentricity. 
This implies that, during every major merger, a stellar mass in the range $0.5-1 M$ is  ejected from the bulge, a possible clue to 
explain the observed mass deficit in ellipticals (Haenhelt \& Kauffmann 2002; Ravindranath, Ho \& Filippenko 2002; 
Volonteri et al. 2003; Graham 2004). 
The properties of the ejected stars are well defined. They typically corotate with the MBHB, 
and show a large degree of flatness, as they are, preferentially, ejected in the binary plane. 
The ejected sub-population, if detected in nearby galaxies, would be an unambiguous sign of a relatively recent major merger. 
Though the total ejected stellar mass does not depend on the exact value of the binary eccentricity, 
the velocity function of the population does.  
In fact, an highly eccentric binary tends to produce a more pronounced tail of hypervelocity stars.    

The relevant quantity to asses coalescence is the ratio $a_f/a_{\rm GW}$, where the final separation $a_f$ is the orbital separation after loss cone depletion. In general, $a_f$ is reached in a short time, $t \sim 10^7$ yrs. 
Though rapid, slingshot is, in general, not sufficient to drive binaries to the gravitational wave emission regime, unless 
the binary is very massive ($M \gtrsim 10^8 \msun$), and/or very eccentric ($e \gtrsim 0.7$), and/or very unequal mass 
($q\lesssim 0.01$, in fact the gap $a_h-a_{\rm GW}$ is lower for unequal mass binaries, though the shrinking factor is small). 
Our conlusion is supported by recent N--body simulations (Makino \& Funato 2004; Berczik et al. 2005).
In terms of {\it LISA} binaries ($10^4-10^7 \msun$), it is then important to study in details the possible role of other 
mechanisms, such as non sphericity of the stellar bulge and presence of circum--nuclear gaseous disks.

\begin{acknowledgements}
I would like to thank the LOC for the organization of a very  exciting workshop. 

\end{acknowledgements}

\bibliographystyle{aa}

\end{document}